# Supersensitive multipurpose/multifunctional avalanche gaseous detectors for environmental, hazard, intrusion systems (SMART)

Marcello Abbrescia[1,2], Giacinto De Cataldo[2,3], Antonio di Mauro[3], Paolo Martinengo[3], Cosimo Pastore[2], Vladimir Peskov[3,4], *Francesco Pietropaolo[3,5], Giacomo Volpe[1,2] and Alexey Rodionov[4,6]

[1]Dipartimento Interateneo di Fisica "M. Merlin", Via G. Amendola, 173, 70126 Bari, Italy; [2]Sezione INFN, Via E. Orabona 4, I-70125 Bari, Italy; [3]European Organization for Nuclear Research (CERN), Esplanade des Particules 1, CH-1217 Meyrin, Switzerland; [4]N.N. Semenov Institute for Chemical Physics, Kesygina street, 4, 117979 Moscow, Russia; [5]INFN Sezione di Padova, Via F. Marzolo, 8, 35131 Padua, Italy; [6]REMOS Electronic company, Programistov street, 4, 141980 Dubna, Russia.
*Corresponding author: Francesco.Pietropaolo@cern.ch

**ABSTRACT**

The aim of this project is to develop a prototype of integrated detector system to monitor environmental hazard: appearance of flames, smoke, sparks or dangerous gases (flammable, toxic, radioactive). We built and successfully tested prototypes of all components of the system. Our sensors are based on established CERN technologies and have superior characteristics, featuring between 10 to 1000 times higher sensitivity than the best commercial sensors. The final version of our device would consist of multifunctional sensors assembled in a single unit; each sensor will perform a specific task and deliver information to a common computing centre via cellular or satellite phone protocols.

## 1. INTRODUCTION

An early detection of hazard environmental conditions and fast automatic issuing of alarm signals to emergency services may save human life and properties, industrial and agricultural infrastructures.

Our Integrated Detector System (IDS) is designed for this scope. It contains innovative detectors of flames, sparks, smoke and dangerous gases, combined with one or more low power pulsed ultraviolet (UV) lamps, assembled in one compact device. By design, all detectors can operate in 100% humid air. The pulsed UV lamps permit continuous long-term operation of the system and, if necessary, can also be used for the activation of a special compact automatic device for cleaning of the air coming to the sensors from dust particles. All these features allow the system to operate in harsh environmental conditions.

IDS can be a part of a network, capable of detecting any of the above stated hazards. The network can be deployed over critical and large areas; it will be self-powered and remotely read-out.

As a result of this work all components of the IDS, based on advanced CERN radiation technology, were built. Tests proved that they all have outstanding technical properties (such as sensitivity, multi-functionality and the ability to operate in harsh environmental conditions) compared to commercial devices.

## 2. STATE OF THE ART

There are many types of detectors of flame, smoke and dangerous gases on the market. They are classified by the sensitivity, the size and the price.

The most sensitive flame detectors are the so called EN-54-10 Class-1 in European Union standards. An example of this type of sensors is the Net Safety UVS-A detector, based on a gas chamber with a metallic photocathode. Most of the commercial flame detectors and are not able to distinguish signals coming from the flame, cosmic radiation or sparks. This limits their applications.

Among sensors of dangerous gases, a special place is taken by the photoionization detectors (PID). They are ionization chambers operating in air and combined with UV lamps operating in continuous mode (named in the following as "continuous lamps") and emitting photons, with energy in the range of $E_v$ = 9.5-11.7 eV. Photons with this energy cannot ionize air. However, if gases with ionization potential $E_i < E_v$ enter the ionization chamber, they will be photo-ionized by the lamp radiation: the generated electron-ion pairs moving in the electric field produce a detectable current. By using continuous UV lamps with different photons energies, one can roughly identify the gases. The sensitivity of this method, for a given lamp intensity and detector fiducial volume, is determined by the sensitivity of the used pico-ammeter.

The ordinary PID ionization chamber can detect ~1 ppm of gases with small ionization potential and the best one down to 0.1 ppm. In commercial PID devices the UV lamps have limited lifetime and for this reason they are switched on only during short-timed measurements.

In the category of dangerous gases falls Radon (Rn). There are many types of Rn detectors in the market, the most advanced among them are gaseous ionization chambers and solid-state sensors which operate in real time/online mode. In both detectors the signals are generated mainly by the motion of primary electrons created by alpha particles, towards the electrodes, where they are finally collected. They have the highest sensitivity but they are incapable of operating in harsh environmental conditions.

## 3. BREAKTHROUGH CHARACTER OF THE PROJECT

There are several important differences of our approach compared to the state of arts. All IDS sensors are based on avalanche detectors developed at CERN and widely used in HEP experiments. However, this technology is not spread yet to industry, placing our detector in privilege position due to their superior characteristics and lower cost with respect to commercial devices.

In our flame sensors, hole-type or wire-type avalanche gaseous detectors, developed at CERN, are combined with high efficiency gaseous or solid photocathodes, for example CsI, and are filled with optimized gas mixtures. As a result, their sensitivity to flames, depending on the design, can be up to 1000 times higher than commercial devices. They operate in proportional mode and have also 1000 times better time resolution; as a consequence, they can distinguish between flames, sparks and cosmic radiation, by means of photon arrival rate and pile-up. In combination with pulse lamps they can detect flames, sparks and smoke simultaneously. Advanced designs of our flame detector have imaging capabilities allowing to visualize flames and sparks and determine their position [1].

In contrast to commercial sensors, our PID detectors are operating in avalanche mode, where the primary electrons are multiplied. In air the avalanche process is peculiar. Our studies showed that primary electrons created by the UV light are immediately attached to the surrounding molecules and form negative ions, which drift toward the multiplication structure, where, in the strong electric field, the electrons are detached and initiate avalanches. Understanding this process permitted to exploit this effect in practical devices and increase their sensitivity. Beside these ionization features, absorption measurements were introduced allowing to better classify the investigated gas. Continuous PID lamps have limited lifetime. For these reasons, we investigated alternative approaches. The continuous monitoring will be performed by a low power pulsed UV lamp, where the lamp is periodically switched on for a short time interval (20 ns - 1 μs, depending on the lamp design and the gas filling) allowing a remarkable increase of the lamp lifetime. If a dangerous gas with $E_i < E_v$ appears in the detector volume, it will be photo-ionized and a large number of primary electrons will be created within the pulse duration and they will trigger avalanches. The avalanche signal can be used for the short-term activation of standard PID lamps allowing to better identify to which group the given gas belongs to.

In the Rn detector, we also explore avalanche multiplication in air, offering much higher signal to noise ratio compared to any commercial devices operating in simple primary charge collection mode. This feature opens the possibility to operate in harsh outdoor conditions: vibrations, electrical pick-ups, elevated temperatures, high humidity.

In IDS, all detectors will interact with each other, adding possibly new features: for example, the signal from the smoke detector may trigger the activation of the gas cleaning device installed in front of Rn detector, making it able to operate in fog and dust environmental conditions.

## 4. PROJECT RESULTS

The goal of SMART project was to develop a concept of the IDS and to build and test its component. Schematics and pictures of some essential parts of the prototype are presented in Fig. 1. The IDS contain detectors of flames, sparks, smokes and dangerous gases as well as one or several pulsed UV lamps working simultaneously.

During the project, most of the time was devoted to build and successfully test all components of the IDS. In order to address in a common framework all the applications mentioned above, two main designs of specially-developed gaseous detectors were studied, a micropattern/microwire-type (MWPC) and a spark-protected resistive gas electron multiplier (RETGEM), combined with solid photocathode, e.g. CsI or CuI. The detailed description of the detector designs, setups and obtained results can be found in the following papers [2,3]. In Ref. [4] there is also a collection of videos showing various tests performed with these detectors. In the following, few examples of some tests performed with flame, dangerous gases and Rn detectors are given.

In Table 1, results of tests of various prototypes of SMART flame sensors are presented [3]. The sensitivity of our detectors is between 20 and 1000 times higher (depending on the photocathode and the detector design) than class-1 detectors. With such a detector a small flame produced by a candle at a distance of 30 m can be easily detected inside fully illuminated buildings. In the same table, the temperature interval within the tests were performed is also shown.

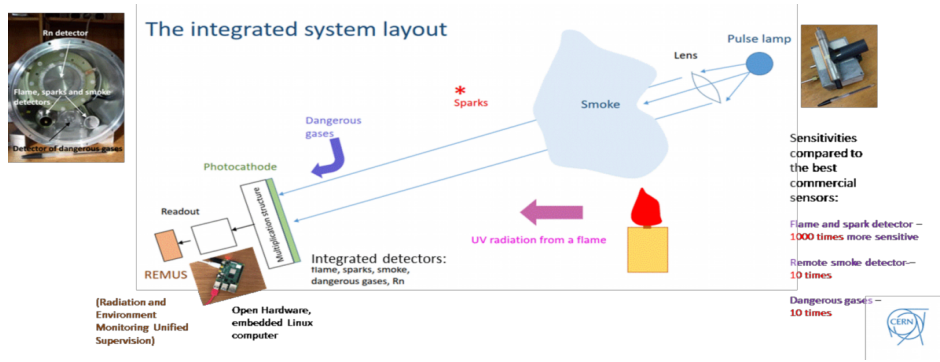

**Fig. 1.** Schematic layout of the IDS. In the insertion, photographs of some of its components are presented.

The important advantage of our detectors is that they can operate in proportional mode, with the signal proportional to the number of primary electrons, $n_0$, simultaneously created by the ionizing radiation in its volume. The value $n_0$ depends on the source of radiation: $n_0 = 1$ in case of flame, while $n_0 \gg 1$ in the case of external sparks and rare cosmic rays. There are several methods allowing to distinguish between sparks appearing in the monitoring area and cosmic radiation, typically using two detectors operating in signal coincidence mode.

We exploited proportional mode to detect simultaneously also flames, sparks and smoke. Such an apparatus is included in Fig 1. It contains a UV pulsed lamp combined with the flame detector. The UV light periodically generates pulses of large amplitudes ($n_0 \gg 1$). At the same time the detector is able to record pulses with $n_0 = 1$, produced by flames (when they happen). If there is smoke (or any other obstacle) on the way from the pulsed UV light, it causes the signal attenuation and the amplitude of the recorded periodical pulses decreases (see Fig. 2); this can be used to generate a smoke alarm signal.

In table 2 preliminary results obtained with IDS PID detectors operating at a gas gain $A = 8$ and 100 are shown. For comparison, data from standard ionization chamber mode ($A = 1$), typically used in commercial devices, are also shown. Two continuous UV light sources were used emitting at $\lambda = 185$ nm and $\lambda = 125$ nm. Although at high gas gains the accuracy of measurements was affected by the electronics noise, it is clearly seen that the sensitivity in determining the vapour concentration increases with gain: roughly a factor close to 5 at gain $A = 8$ and more than 10 at higher gains.

Preliminary tests of the IDS MWPC Rn detectors indicate that they have efficiencies for alpha particles approaching 100% (Fig. 3), low counting rate spurious pulses (well below 1 Hz), can operate stably in 100% humid air and at elevated temperatures. We designed and tested special automatic cleaning system, based on corona discharge, which will allow the Rn detector to operate in fog and dust conditions. Systematic measurements of its sensitivity to Rn are currently in preparation, however early measurements, indicate that it is equal to that of the best commercial devices [5]. The general conclusion is that our detectors of flames, sparks, smoke and dangerous gases fulfil the main requirements of the proposed IDS: they have high sensitivity, high time resolution and able to operate in outdoor conditions.

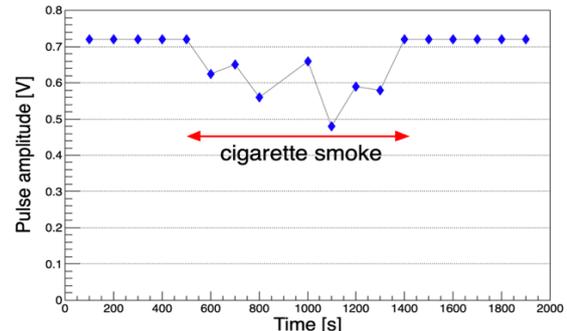

**Fig.2.** Typical changes in amplitudes of periodical pulses caused by the sudden smoke from a cigarette, crossing the lamp beam.

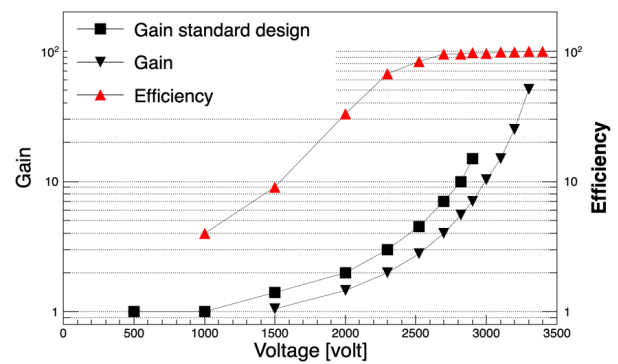

**Fig. 3.** Gain and efficiency as a function of the applied voltage for the Rn detector based on the MWPC, modified to operate in 100% humid air. For comparison the gain of a standard MWPC is also shown.

**Tab. 1.** Relative sensitivities (with respect to class-1 detector) of SMART flame detectors, based on RETGEM, or MWPC combined with CsI, CuI and Ni photocathodes, able to operate stably in the indicated temperature intervals. Accuracy of the measurements is ~5%.

| Detector type | Single Wire | Single Wire | RETGEM | MWPC | MWPC |
|---|---|---|---|---|---|
| *Photocathode* | CsI | CsI + filter | CuI | CuI | Ni |
| *Rel. Sensitivity* | 1140 | 230 | 110 | 28 | 22 |
| *Temperature Interval (°C)* | [-100, +60] | [-100, +60] | [-65, +90] | [-65, +90] | [-65, +90] |
| *Solar blindness* | No, in direct sunlight | yes | yes | yes | yes |
| *Application* | Indoor | Outdoor/forest fire | Outdoor | Outdoor | Outdoor |

**Tab. 2.** Values of the lowest concentrations of some vapours in air that is possible to measure in ionization chamber mode ($A = 1$) and in avalanche mode at gain $A = 8$ and 100 with two monochromatic light sources, emitting at $\lambda = 185$ and 125 nm. Accuracy of measurements at gain 1 is ~10% at gain 8 is ~20% and at gain 100 is ~35% due to high noise of floating pico-ammeters. For this reason, at gain $A = 100$ measurements were performed with a standard pico-ammeter (accuracy of about 15%).

| Vapours<br>Sensor type, wave length | | Ethylferrovene,<br>Single Wire, 185 nm | Ethylferrovene,<br>Single Wire, 125 nm | Ethylferrovene,<br>MWPC, 125 nm | Benzene<br>MWPC, 125 nm | Acetone,<br>MWPC, 125 nm |
|---|---|---|---|---|---|---|
| Lowest concentration (ppm) | *gain 1* | 2.4 | 0.5 | 0.16 | 0.3 | 0.15 |
| | *gain 8* | 0.4 | 0.07 | 0.03 | 0.06 | 0.035 |
| | *gain 100* | 0.03 | | | | |

## 5. FUTURE PROJECT VISION

During ATTRACT-1, we performed basic technology research on the concept of an IDS for environmental monitoring purpose and prove its feasibility by means of experimental laboratory tests showing that its components meet all specific requirements.

If we will be selected for ATTRACT-2, this will give us the possibility to build and test in real indoor and outdoor conditions, an industrial prototype of a fully automatic, intelligent and multifunctional device for monitoring hazard environmental conditions able to issue early warning signals to fire brigade and other emergency services. Below concrete steps in this direction are described and required resources are indicated.

### 5.1. Technology Scaling

The first step will be to create a consortium of industrial partners having experience in developing, producing and selling similar sensors. We already established connections with several companies in Europe and elsewhere to which our detectors were shown in operation and they were convinced of the advantages of our technology. One of the initial tasks of the consortium will be to professionally evaluate market potentials of the IDS and create a road map on how to proceed from the existing laboratory prototype to a commercial device.

The second step will be to convert the existing simplified laboratory prototype of the IDS to a commercial prototype. It will be remotely connected to a computer centre. This step will have some technological challenges in order to make the devices reliable in operation and to ensure its low power consumption and competitive cost. This commercial prototype can be used for demonstrations to the investors, partners and journalists or to other potential customers in order to convince them of the viability of the chosen approach.

The aim of the third step will be to confirm that the technology components can be integrated into a complete system whose performance and operation are defined by the anticipated operating scenarios. This will include all necessary tests proving that the system is ready for commercialization, e.g., false signals rejection, resistance to shocks, vibrations and electrical pick-ups.

### 5.2. Project Synergies and Outreach

At present our collaboration consists of universities, research institutes and commercial partners –companies having experience in developing flame and dangerous gases sensors. Several other companies also expressed an interest in our technology, and we are planning to include them into the consortium of SMART-2 project. We expect that due to the combined effect of research and business expertise our reinforced consortium will be capable achieving TRL 5-7 level.

With one of our partners (REMOS) we are planning to continue efforts to reinforce our system by including hyper spectroscopy [6,7] capability allowing to monitor from drones or helicopters the chemical composition and the level of pollutions on the earth or water surfaces. We are also currently developing a detector which will allow to extend hyper spectroscopic method to the UV region (185-260 nm) by using artificial UV sources installed on the flying carriers. This will allow to strongly increase the detection power of pollution's chemical composition. We consider clustering this part of our activity with other ATTRACT projects dealing with hyper spectroscopy.

Our activity will be publicly spread via demonstrators of our technology at public events, exhibitions, and through publication and specially organized web pages.

### 5.3. Technology application and demonstration cases

The main task of the IDS is to identify the environmental hazards and to issue an early warning signals to emergency teams. Some components of our system can be also used individually to monitor the specific hazard situation. For example, our supersensitive detectors of flame can be deployed on drones or in strategic positions for monitoring flame hazard situation in forests and bushes.

Sparks, flames and smoke detectors can be also used in other potentially dangerous areas: plants, factories, chemical product storage houses, oil platforms, etc.

A separate task can be performed by our Rn detectors. Some recent studies have shown the possibility to correlate rapid changes of the Radon concentration in soil or in groundwater to the early prediction of earthquakes; In order to verify such observations on a more solid statistical ground, it is needed to create a wide network of cheap, compact and high sensitivity Rn detectors, deployed in key points where earthquakes may potentially occur. The existing excellent commercial detectors are too expensive to be employed in large numbers, while the sensors developed in the framework of SMART project are characterized by higher simplicity, low costs and capability of operation in outdoor environmental conditions.

### 5.4. Technology commercialization

During the last stage of the SMART-1, in collaboration with Fenno-Aurum we already developed and produced commercial prototypes of flame, spark and smoke detectors. A commercial prototype of the detector for flame visualization, suitable also for the UV hyper spectroscopy, was built with REMOS and will be tested in September this year. During ATTRACT-2, we plan to create our own setup for manufacturing commercial prototypes of all components of the IDS. A special task will be to develop a dedicated read-out electronics and compact power suppliers in order to make it fully automatic, autonomous and with low power consumption, to ensure reliable network communication. We are already in contact with some companies and the work in this direction is in progress. Finally, all necessary tests in indoor and outdoor conditions should be done, to validate the results. The estimation of the required budget is presented in Table 3 (in kEuros).

### 5.5. Envisioned risks

Risks may come from new and more cost-efficient competitive technologies, if they appear during the time period of ATTRACT-2. By a careful follow up of the technology evolution we can correspondingly improve or modify our design.

**Tab. 3.** Estimated budget for ATTRACT-2

| Year | 1 | 2 | 3 |
|---|---|---|---|
| Tasks | Equipment purchase for IDS industrial prototype (**100**) Mechanical and electronics works (**20**) Salaries (**150**) Travels (**10**) | Setup for IDS production-(**50**) Electronics and mechanical works (**20**) Salaries (**150**) Travels (**10**) | Manufacturing and final tests of a commercial IDS system (**150**) Salaries (**150**) Travels (**20**) |
| **Budget** | **280** | **230** | **320** |

### 5.6. Liaison with Student Teams and Socio-Economic Study

During ATTRACT-1, our team had several interactions with MSc level student teams from institutes, such as the High Tech XL (https://www.hightechxl.com/), the CERN Entrepreneurship Student Programme (CESP), and Norwegian University of Science and Technology, School of Entrepreneurship (NTNU), all interested in applications of our flame detectors.

We lead the experimental work of a student team from Alto University, Finland, aiming at preparation and installation of hardware for our fame detectors on drones from early detection of forest fires. The flight test was delayed due to the COVID-19.

In the framework of ATTRACT-2 we plan to contribute with all necessary information about our integration system, its performance and applications to socio-economic advertisement and ecosystem.

### 6. ACKNOWLEDGEMENT

We warmly thank M. Meli, J. Ven Bellevin and M. Van Stenis for the technical support. We are also grateful to the CERN RD 51 lab and the CERN Neutrino Platform group for helping us with equipment and expertise. This project has received funding from the ATTRACT project funded by the EC under Grant Agreement 777222.